\documentstyle[12pt]{ioplppt}

\begin{document}

\sloppy

\jl{6}

\paper{ 
Can gravity appear due to polarization of instantons in the 
${\bf SO(4)}$ gauge theory?}[Can gravity appear due to ... instantons ...?]

\author{M. Yu. Kuchiev \ftnote{1}{E-mail: kuchiev@newt.phys.unsw.edu.au} }

\address{School of Physics, University of New South Wales,
Sydney, 2052, Australia}

\begin{abstract}
Conventional non-Abelian SO(4)
gauge theory is  able 
to describe gravity provided the gauge field 
possesses a specific polarized vacuum
state. In this vacuum the instantons and antiinstantons have a
preferred direction of orientation. Their orientation
plays the role of the order parameter for the 
polarized phase of the  gauge field.
The interaction of a  weak and smooth gauge field
with the polarized vacuum is described by an effective 
long-range action which is identical to the 
Hilbert action of general relativity.
In the classical limit this action results in 
the Einstein equations of general relativity.
Gravitational waves appear as the 
mode describing propagation of the gauge field which strongly 
interacts with the oriented  instantons.
The Newton gravitational constant describes the
density of the considered phase of the gauge field. The 
radius of the instantons under consideration is comparable with the 
Planck radius.
\end{abstract}


\date{\today}

\maketitle

\section{Introduction} 
\label{int}

I wish to present arguments favoring a scenario in
which gravity can arise as a particular effect 
in the framework of the conventional 
Yang-Mills gauge theory \cite{YM}
formulated in flat space-time.
The idea was first reported in \cite{IAP}, this 
paper presents it in detail.
We will suppose that the Lagrangian of the theory 
describes the usual degrees of freedom 
of gauge theory, i.e.
gauge bosons  interacting with 
fermions and scalars,
while there are no geometrical degrees of freedom 
on the Lagrangian level.
In particular, the Lagrangian does not contain gravitons.
Our goal is to show that all  
geometrical objects 
necessary for the gravitational phenomenon
can originate from the gauge degrees of freedom
if a particular nontrivial vacuum
state, which we will call the vacuum with polarized instantons
develops in the $SO(4)$ gauge theory.

Let us  first describe 
the main idea in  general terms. 
To do this let us imagine that 
in some  conventional gauge theory 
there develops a particular 
nontrivial vacuum state.
Its nature is discussed below in detail.
At the moment let us only assume that
this this nontrivial vacuum has a very strong impact on
the properties of excitations  
of the gauge field.
Instead of the usual 
spin-1 gauge boson there appears quite new 
massless  spin-2 excitation.
Certainly this is a very peculiar property, 
but if it exists, it provides a  possibility to identify
this massless  mode with the  graviton. 
Furthermore one can hope to evaluate all manifestations
of gravity from  an effective theory which  should describe
propagation of this low-energy mode.
In the classical limit
one can hope also that this effective theory should reproduce the 
Einstein equations of general relativity. 
Developing this line of reasoning
one should keep in mind a sufficiently long list of important 
relevant questions.
The most urgent ones are listed below.\\
1.In which gauge group one should look for the effect?\\
2.What is the origin of the necessary nontrivial
vacuum,  roughly speaking from which fields
should it be constructed?\\
3.What is  the nature of the order parameter
which governs  the nontrivial phase?  In particular,
which symmetry should  it possess?\\
4.What is the nature of the low-energy excitations in this vacuum?
As  said above, this excitation should be
the massless spin-2 mode.\\
5.What is the effective low-energy theory governing classical 
propagation of this massless spin-2 mode?\\
6. How important  are the
quantum corrections? One should hope 
that they neither confine the spin-2 mode,
nor supply it with the mass in the infrared region.\\
7. What is the origin of the phase transition
into the necessary vacuum state?\\
8. Which   phase transitions separate the
necessary vacuum from  conventional vacua of gauge theory?

All these questions  are concentrating around  the two
major problems. 
One of them  is the existence of the
necessary vacuum. 
Possible models in which it can arise
have been considered 
in recent Refs.\cite{det,PRdet}.
The other problem is 
physical manifestations of the nontrivial vacuum,
granted it exists. This later problem  is in the focus of this paper. 

Several  reasons make this study necessary.
Firstly,  the physical
nature of the excitations proves to be very interesting.
Secondly, one should know precisely the properties of the 
excitations in order to construct and tune a model in which
the necessary ground state  appears.
Thirdly, even in such a long known phase as the QCD vacuum
the properties of the vacuum state are not fully understood
up to now. Instantons  are known to play an important
role there, but verification of this fact remains
an interesting as well as nontrivial problem \cite{shu}. 
This paper proposes to
consider a new phase in gauge theory
in which instantons play an important role.
It could also prove to be sufficiently complicated.
To begin such a study one should fully
realize its purpose.
Having these arguments in mind I address in this paper
some questions from the above list
in detail, while only qualitatively discussing others.

The idea discussed can suggest
a novel approach to the old problem of quantization
of gravity having several important  advantages.
One of them is the renormalizability which  is guaranteed
from the very beginning
because on the basic Lagrangian level the theory 
is the usual renormalizable gauge theory. 
It is important also that the basic 
idea is sufficiently  simple being formulated by means
of a pure field theory.  It needs the single
tool, the nontrivial condensate, to explain the
single phenomenon, gravity. 
The string theory  \cite{GrS}, which
for a long period of time has been considered as one of the most
advanced candidates for description of quantum gravity
offers much more complicated approach.

One of the popular modern developments in
quantum gravity was originated   in \cite{ash},
where it was argued that the variables of gauge theory 
can be  used for description of  geometrical objects
of the Riemann geometry. 
In this approach it is assumed that
the geometrical degrees of freedom do exist on the basic Lagrangian level.
In contrast, in this paper 
we suppose that geometry does not manifest
itself on the Lagrangian level, 
providing  only an effective description 
for some particular low energy  degrees of freedom of gauge field.

Let us now  describe shortly the answers 
which this paper  proposes  to some of the listed above questions.
First, the  gauge group has to be $SO(4)$, or bigger one.
Furthermore, the phase considered may be described 
in terms of nontrivial topological excitations of the gauge
field. In this sense the condensate 
is composed of the $SO(4)$ gauge field.
The necessary phase
can be described in terms of 
the  BPST instanton \cite{bpst}.
An instanton is known to possess eight degrees
of freedom. Four of them give its position, 
one is its radius. The rest three 
describe its orientation
which plays a crucial 
role in our consideration. 
In the usual phases of gauge theory, for example in the QCD vacuum
orientations of instantons are arbitrary.
In this paper  we discuss a phase in which
instantons  have a preferred direction
of orientation. This means that a mean value of 
 orientation is nonzero.
A possible way to visualize
this vacuum, to find  a simple physical analogy, 
is to compare it with the  usual ferromagnetic 
or antiferromagnetic phases in condensed matter physics
in which spins of atoms or electrons have preferred orientation.
The instantons constituting
this vacuum will be called ``polarized instantons'' 
or ``the condensate of polarized instantons''. 
The vacuum  itself will be referred to 
as ``the polarized vacuum''.
The density of the condensate of polarized instantons
is described by  a length parameter which depends on radiuses
and separations of the polarized instantons.
This length parameter proves to be  equal to the Planck radius, 
which is only possible if
radiuses of the considered instantons are
comparable with the Planck radius.
The existence of this length parameter permits 
the Newton gravitational constant to appear in the theory
as the inverse density of the condensate.
Thus the main gravitational parameter  absent in the initial 
Lagrangian is introduced into the theory by
the nontrivial vacuum state. 
In Section \ref{res} we will return to the questions listed above 
and discuss them in more detail.

It is necessary to mention that
the existence of a state with polarized instantons 
does not  come into contradiction  
with a general principle of gauge invariance
which in  the context considered is 
known as the Elitzur theorem \cite{eli}, see also  
Refs.\cite{pol,ham}.
The  theorem  forbids  existence of any vacuum state 
in which  local gauge invariance
is broken spontaneously by a
non-gauge-invariant  mean value of a field.
The orientation of an instanton is a gauge invariant
parameter. Therefore 
polarization of instantons does not break spontaneously 
gauge invariance, and the vacuum state
with polarized instantons is not forbidden.

We will mainly restrict our  discussion to 
the dilute gas approximation for instantons, mainly to secure  
simplicity of consideration.
In this approximations the 
gases of instantons and antiinstantons can coexist.
We will suppose that both gases of instantons and antiinstantons
are polarized.
The mean values of 
matrixes which describe orientations of polarized instantons and 
antiinstantons play the  role of the order parameter
for the considered   phase.
It  will be argued that this order parameter
should have a particular symmetry. 
Namely 
orientations of all polarized topological excitations should be described
by one $6\times 6 $ matrix which belongs to $SO(3,3)$.

The main result of this paper
can be formulated  as the following statement.
Suppose that instantons and antiinstantons 
in the vacuum state of the $SO(4)$ gauge theory
have preferred orientations which are described
by a matrix from the $SO(3,3)$ group.
Then long-wave excitations above this vacuum 
are the spin-2 massless particles which on the classical
level are described by the Hilbert action
and the Einstein equations of general relativity.

\section{Instantons in the external field} 
\label{pair}
Consider Euclidean formulation of the  $SO(4)$ gauge theory.
The gauge algebra for $SO(4)$ gauge group consists of  two $su(2)$ 
gauge subalgebras, $so(4) = su(2)+su(2)$. 
The  instantons and antiinstantons  can belong to any one
of these two available $su(2)$ gauge subalgebras.
It is convenient to  choose the generators for one 
$su(2)$ gauge subalgebra to be $(- 1/2)\eta^{aij}$ and the 
generators for the other  one
to be $(-1/2) \bar \eta^{aij}$. 
To distinguish between these two subalgebras we will  refer to  them as 
$su(2)\eta$ and $su(2) \bar \eta$. 
Symbols $\eta^{aij}, \bar \eta^{aij}$
are the usual 't Hooft symbols, $a =1,2,3,~ i,j = 1, \cdots,4$ which
give a  full set of  $4\times 4$ matrixes antisymmetric in $ij$  \cite{tH}. 
In this notation
the gauge potential and the strength of the gauge field are
\begin{eqnarray} \label{Af}
A^{ij}_{\mu}&=&-\frac{1}{2} (A^a_{\mu} \eta^{aij} + 
\bar A^a_{\mu} \bar \eta^{aij}), \\ \label{F}
F^{ij}_{\mu \nu}&=&-\frac{1}{2} (F^a_{\mu \nu} \eta^{aij} + 
\bar F^a_{\mu \nu} \bar \eta^{aij}),
\end{eqnarray}
where $A^a_{\mu}$ and $F^a_{\mu \nu}$ belong to $su(2) \eta$ and 
$\bar A^a_\mu,\bar F^a_{\mu\nu}$ belong to $su(2) \bar \eta $.
The Yang-Mills action reads
\begin{equation}\label{yma}
S= \frac{1}{4g^2}\int F^{ij}_{\mu\nu}(x)F^{ij}_{\mu\nu}(x) \, d^4x.
\end{equation}
The Latin indexes $i,j=1,.\cdots,4$ label the 
isotopic indexes, while the Greek indexes $\mu,\nu=1,\cdots,4$
label the indexes in Euclidean coordinate space.
Remember that we consider the usual 
gauge field theory in  flat space-time.
For the chosen normalization of generators  
the relation between the gauge potential and the field strength reads
\begin{equation} \label{FA}
F_{\mu\nu}^{ij}=\partial_\mu A_{\nu}^{ij}-
\partial_\nu A_{\mu}^{ij}
+A_{\mu}^{ik}A_{\nu}^{kj}- A_{\nu}^{ik}A_{\mu}^{kj}.
\end{equation}

For our purposes it is  important to consider
an interaction of nontrivial topological excitations, 
instantons and antiinstantons
with an external gauge field which has trivial topological structure. 
Consider first a single instanton in an
external gauge field.  It is sufficient to assume
that this external field is weak and smooth,  i.e.
it is weaker than the field of the instanton inside the instanton
and varies much more smoothly than the instanton field which
sharply decreases outside the instanton radius. 
Thus formulated problem was  first considered 
by Callan, Dashen and Gross in \cite{CDG} where
it was shown 
 that the interaction of an instanton 
with the external field is described by an effective action
\begin{equation}\label{cdg}
\Delta S = 
\frac{2 \pi^2 \rho^2}{g^2}\bar \eta^{a\mu\nu} D^{ab}F^b_{\mu\nu}(x_0).
\end{equation}
Here $\rho$ and $x_0$ are the radius and position of the instanton.
The matrix  $D^{ab} \in SO(3)$ describes the orientation of the instanton 
in the $su(2)$ gauge subalgebra where the instanton 
belongs (suppose for example that it is $su(2)\eta$).
Its definition is given in Appendix A, see (\ref{CU}).
$F^b_{\mu\nu}(x)$
is the gauge field in the subalgebra where the instanton belongs.
This field has to be taken in the singular gauge \cite{tH}.

The  interaction of an antiinstanton with an external field
is described similarly. The only distinction
is that it produces the corresponding term with the 't Hooft symbol 
$\eta^{a\mu\nu}$ instead of $\bar\eta^{a\mu\nu}$  which stands in
(\ref{cdg}). Notice that we use 
$\eta^{a\mu\nu}, \bar \eta^{a\mu\nu}$ as generators of rotations in
the coordinate space, while
$\eta^{aij},\bar \eta^{aij}$ generate rotations in the isotopic space.
These two sets of symbols are defined in different spaces,
but numerically they are identical of course. 

It is worth to stress
several interesting and important properties of the action 
(\ref{cdg}). First of all
it has a very unusual structure being
linear in the external field. At first sight
this fact looks as a paradox because
an instanton provides a minimum for the action and
one could anticipate  only quadratic terms in the weak field to appear.
The paradox is resolved
by the fact that the linear term arises from the region of large separations
from the instanton
where  the external field 
exceeds the instanton field, resulting in breaking down of 
the naive perturbation theory.
This question is discussed in detail in Appendix \ref{A}
which develops the approach of \cite{CDG} 
to cover the case of several overlapping instantons.
The action (\ref{cdg}) is derived 
starting from the usual Yang-Mills action
in which the external field is considered as a perturbation.
The contribution to the
action linear in the external field 
 is found as an integral over a sphere which is well 
separated from instantons. 
An alternative way to derive the action (\ref{cdg})
was suggested in \cite{VZNS} where an
instanton was  considered as an effective source for 
gauge bosons. In this approach the linear
term in the action arises due to the large probability
of  creating small-momenta bosons by an instanton.

Furthermore,
it is important to stress 
that the orientation of an instanton is
a gauge invariant parameter. A  convenient way to 
define the orientation is given by the general n-instanton
ADHM solution of Ref.\cite{ADHM}, for a review see
\cite{pras}. This solution  is briefly discussed
in Appendix A where Eq.(\ref{CU}) defines the 
matrix of the instanton orientation $D^{ab}$ in terms of
variables of the ADHM construction.
The fact that the orientation  of an instanton
is a gauge invariant parameter is closely connected
with gauge invariance of the action (\ref{cdg}). This action 
is derived from the usual Yang-Mills action, 
from which it inherits gauge invariance. The invariance is
guaranteed by the mentioned explicit requirement  that the
field $F^a_{\mu\nu}(x_0)$
has to be considered in the singular gauge.
This  fixing of the gauge indicates that
there is no spontaneous breaking of the local symmetry.

The last comment addresses the behaviour of several
instantons.
It is found in Appendix \ref{A}  that they
interact with the external field as individual
objects. This statement is trivial  when the dilute
gas approximation is valid, but surprisingly
it stands for any separation between instantons,
even if they strongly overlap. 
In this paper we will not push this argument further
on restricting our consideration by the dilute gas
approximation. Still the  independent interaction
of instantons with the external field gives a 
hope that  final conclusions of the paper
can be more reliable than the dilute gas approximation
which is used to derive them.

The discussion  
of the action (\ref{cdg}) given above 
permits us now to
address the following important for us
problem. Suppose that there
is a number of instantons and antiinstantons which
belong to either $su(2)\eta$ or $su(2)\bar\eta$ gauge
algebras and  satisfy the dilute gas approximation.
Suppose also that there is  an 
additional gauge  field $F^{ij}_{\mu\nu}(x)$ 
which  has a trivial topological
structure, is weaker than the fields of the instantons 
and varies smoothly
for  the distances which characterize separations and radiuses
of instantons.
In this situation we immediately deduce 
that the influence of the external field 
on  instantons and antiinstantons
can be described by
the effective action which is equal to the sum
of terms  of the type of (\ref{cdg}) which describe 
independent interaction of instantons and antiinstantons
with the field.
Using definition (\ref{F}) one can present 
the corresponding action in the following form
\begin{equation} \label{4gas}
\Delta S = -\frac{\pi^2}{g^2} \sum_k
\eta^{A\mu\nu}\eta^{Bij}  
T^{AB}_k \rho^2_k F^{ij}_{\mu\nu}(x_k).
\end{equation}
Here an index $k$ runs over all available instantons and antiinstantons
which have radiuses and coordinates $\rho_k$ and $x_k$.
To simplify notation
the 't Hooft symbols are enumerated 
as 6-vectors $\eta^A = (\eta^a,\bar \eta^b),
~A=1,\cdots,6;~
a,b=1,2,3$. More precisely this definition means that
\begin{eqnarray}\label{etaA}
\eta^{Aij} &=& \eta^{aij},~~
\eta^{A\mu\nu} = \eta^{a\mu\nu},~~~{\rm if}~~A=a=1,2,3,
\\ \nonumber
\eta^{Aij} &=& \bar \eta^{bij},~~
\eta^{A\mu\nu} = \bar \eta^{b\mu\nu},
~~~ {\rm if }~~A-3=b=1,2,3.
\end{eqnarray} 
To describe an orientation of every (anti)instanton
it is convenient to introduce the $6\times 6$ matrix $T^{AB}_k,~~
A,B=1,\cdots,6$ 
\begin{equation} \label{CAB}
T_k^{AB} \equiv T_k =
\left(  \begin{array}{cc} C_k &  D_k \\ 
 \bar D_k &   \bar C_k \end{array} \right)
\end{equation}
as a set of four $3\times 3$ matrixes  $C_k,\bar C_k,D_k,\bar D_k$.
For any given $k$-th (anti)instanton only one of these four matrixes 
is essential while the other three are
equal to zero.  This nonzero matrix  belongs to $SO(3)$ and  describes the 
orientation of the $k$-th  topological object in the gauge algebra where
it belongs.
For example, if the $k$-th object is 
an  antiinstanton  in the  $su(2)\eta$ gauge subalgebra, 
then we assume that $C_k\in SO(3)$ describes its orientation 
in the  $su(2)\eta$ while $\bar C_k,D_k,\bar D_k =0$. 
Similarly, $D_k$ 
describes the orientation if the $k$-th object is
an instanton  $\in su(2)\eta$ ($C_k,D_k,\bar D_k =0$), 
$\bar D_k$ describes the orientation  if  the 
antiinstanton  $\in su(2)\bar \eta$ is considered
($C_k,\bar C_k,\bar D_k =0$), and $\bar C_k$ describes
orientation if  the instanton $\in su(2)\bar\eta$ is considered
($C_k,D_k,\bar D_k =0$) .
These definitions can be presented in a short  symbolic form
\begin{equation}\label{T}
T_k = 
\left(  \begin{array}{cc} {\rm antiinstanton} \in su(2)\eta  &  
                        {\rm instanton}  \in su(2)\eta \\ 
{\rm antiinstanton}  \in su(2)\bar \eta &  {\rm instanton} 
\in su(2)\bar\eta \end{array} \right).
\end{equation}
Using definitions (\ref{etaA}),(\ref{CAB})
as well as Eq.(\ref{F}) it is easy to verify that
every term in Eq.(\ref{4gas})   
describes an interaction of some (anti)instanton with
the external field  in accordance with (\ref{cdg}).

Let us address now the question of the behavior of the
ensemble of instantons in the vacuum state assuming that
there exists  a weak and smooth topologically trivial gauge field
$F^{ij}_{\mu\nu}(x)$. One can derive the
effective action which describes 
interaction of the vacuum with this field
averaging Eq.(\ref{4gas}) over
short-range quantum fluctuations in the vacuum.
The result can be presented as  the effective action
\begin{equation} \label{f/4}
\Delta S = 
-\int \eta^{A\mu\nu}\eta^{Aij}{\cal M}^{AB}(x)F^{ij}_{\mu\nu}(x)
\,d^4 x.
\end{equation}
where the matrix ${\cal M}^{AB}(x)$ is
\begin{equation} \label{M^AB}
{\cal M}^{AB}(x) = 
\pi^2 \langle \,
\frac{1}{g^2} 
\rho^2  T^{AB}  n(\rho, T,x)\,\rangle.
\end{equation}
The brackets $\langle\, \rangle$ here describe 
averaging over quantum fluctuations whose
wavelength is shorter than a typical distance
describing variation of the external field.
These fluctuations in the dilute gas approximation for instantons 
should include averaging over positions,
radiuses and orientations of instantons. 
In Eq.(\ref{M^AB}) $ n(\rho, T,x)$ is the concentration
of (anti)instantons which have 
the radius $\rho$ and the orientation described
by the matrix $T\equiv T^{AB}$. In the usual vacuum 
states the concentration of instantons does not depend on
the  orientation, $n(\rho, T,x) \equiv n(\rho, x)$. In that case
an averaging over orientations gives the trivial result
${\cal M}^{AB}(x) \equiv 0$, as mentioned in \cite{VZNS}. 
The  main goal of this paper is to investigate
what happens if the concentration $n(\rho, T,x) $ does depend on the 
orientation $T=T^{AB}$ providing the nonzero value for the matrix 
${\cal M}^{AB}(x)$. 

From (\ref{M^AB}) one can anticipate
that the consequences should be interesting  because
there appears the unusual term linear in the gauge field  in the 
action. To clarify this point   
we are to be more specific about properties of 
the matrix ${\cal M}^{AB}(x)$ assuming that it satisfies
the following three conditions. 
First we will assume that
it is a nondegenarate matrix with positive determinant.
It is convenient to present this assumption  as a statement
that  the matrix ${\cal M}(x)$ is proportional to the unimodular
$6\times 6$ matrix $M(x)$
\begin{equation} \label{M=fM}
{\cal M}^{AB}(x)  = \frac{1}{4}\,f\, M^{AB}(x),~~~~\det M(x) =1.
\end{equation}
Second, we suppose that $f$ introduced above is a positive constant 
\begin{equation}\label{f>0}
f = const > 0.
\end{equation} 
This means  that  the determinant $\det {\cal M}(x) = (f/4)^6$
is a constant. 
Third, we postulate that the matrix $M^{AB}(x)$ possesses a
particular symmetry property, namely that it belongs
to the $SO(3,3)$ group, 
\begin{equation} \label{so33}
M^{AB}(x)\in SO_+(3,3).
\end{equation}
This means that $M=M^{AB}(x)$ 
should satisfy identity
\begin{equation} \label{MSM}
 M \Sigma M^T=
\Sigma.
\end{equation}
Here the matrix $\Sigma^{AB}$ is defined as
\begin{equation}\label{Sig}
\Sigma  =
\left(  \begin{array}{cr} \hat 1 &  \hat 0 \\ \hat 0 & -\hat 1
\end{array} \right),
\end{equation}
where numbers with hats represent $3\times 3$ diagonal
matrixes. 
Notation $SO_+(3,3)$ is used in (\ref{so33})  
to distinguish the subset  of matrixes 
$M\in SO(3,3)$
which can be transformed into the unity matrix ${\bf 1}$
by a continuous function in $SO(3,3)$, 
see Appendix \ref{B}.

Conditions  (\ref{M=fM}),(\ref{f>0}),(\ref{so33})
are to be considered as 
the main {\it assumptions} made in this paper
about   properties of
the vacuum state of the $SO(4)$ gauge theory.
Now we can  clarify the meaning of
the term  ``polarization of instantons'' which was introduced
above intuitively.
We  say that instantons in the vacuum state
are polarized  if the matrix ${\cal M}^{AB}$ 
  which 
defines the mean values
of orientations of (anti)instantons in (\ref{M^AB}) is nonzero.
We will say that the polarization of instantons
has the $SO(3,3)$ symmetry, 
or equivalently there is the  $SO(3,3)$ 
polarization of instantons,  if
Eqs.(\ref{M=fM}),(\ref{f>0}),(\ref{so33}) 
are valid. 
Let us mention once more that the orientation of an instanton
is a gauge invariant parameter. Therefore its 
nonzero mean value does not come into contradiction with
the Elitzur theorem \cite{eli}.

In the following consideration we postulate
that there is  the $SO(3,3)$ polarization of instantons in the vacuum.
This assumption  at the moment looks rather bizarre,
but later development will show its advantages.
Firstly,
in the next Section \ref{eineq} we will verify
that it makes excitations above the vacuum  identical
to gravitational waves. 
This is exactly what we are looking for.
Then in Section \ref{res} we discuss 
what is known about a  way to
justify   assumptions 
(\ref{M=fM}),(\ref{f>0}),(\ref{so33}) 
in the framework of gauge theory.

In the usual  phases  of gauge theory
the mean values of matrixes describing orientations
of (anti)instantons vanish. Therefore one can interpret 
Eqs.(\ref{M=fM}),(\ref{f>0}),(\ref{so33})
as the statement that there exists the new nontrivial
phase of the $SO(4)$ gauge theory. The matrix
$M(x)$ plays the role of the order parameter for this phase.
The (anti)instantons which contribute to the 
nontrivial value of the matrix $T^{AB}$ in (\ref{M^AB}) can
be looked at as a specific condensate.
The constant $f$ characterizes the density of this condensate.
The dimension cm$^{-2}$ of this constant supplies the theory
with the dimensional parameter which later on
 will be interpreted as
the Newton gravitational constant,
see Eq.(\ref{newt}).

In order to examine the consequences of  assumptions 
(\ref{M=fM}),(\ref{f>0}),(\ref{so33}) it is very
instructive to re-write (\ref{f/4}) in another form.
With this purpose let us take into consideration
that there exists
a relation between matrixes belonging 
to $SO(3,3)$ and matrixes
belonging to $SL(4)$ groups.
It states that 
for any $M^{AB} \in SO_+(3,3),~A,B=1,\cdots,6$
there exists some matrix $H^{i\mu}\in SL(4),~i,\mu=1,\cdots,4$ satisfying
\begin{equation} \label{hom}
H^{i\mu}H^{j\nu}-H^{i\nu}H^{j\mu} =
\frac{1}{2} \eta^{A\mu\nu}\eta^{Bij}{M}^{AB}.
\end{equation}
This matrix if defined uniquely up
to a sign factor $\pm H^{i\mu}$. 
The reversed statement
is also true: for any given $H^{i\mu}\in SL(4)$ 
Eq.(\ref{hom}) defines 
the matrix $M^{AB}$ which belongs to $SO_+(3,3)$.
Moreover, it can be shown that (\ref{hom})
gives an isomorphism $SL(4)/(\pm {\bf 1})\equiv SO_+(3,3)\,$.
All these statements are verified in Appendix \ref{B}.

Identifying $M^{AB}(x)=M^{AB}$ one finds from (\ref{hom}) 
$H^{i\mu}=H^{i\mu}(x)\in SL(4)$.
This new matrix is well defined by the matrix   $M^{AB}(x)$
and hence  
can be considered as another representation of the
order parameter for polarized instantons.
Substituting  $H^{i\mu}(x)$ defined by
 (\ref{hom}) into (\ref{M=fM}) one finds that 
the action (\ref{f/4})  
can be rewritten in the following  useful form
\begin{equation} \label{HHF}
\Delta S = -f \int H^{i\mu}(x)H^{j\nu}(x)F^{ij}_{\mu\nu}(x)
\,d^4x.
\end{equation}
Up to now our consideration was fulfilled in the 
orthogonal coordinates which describe the
considered  flat space-time. In these coordinates the 
Lagrangian of the gauge theory is formulated.
It is instructive however to present the action 
(\ref{HHF}) in arbitrary coordinates. 
Under the coordinate transformation
$x^\mu \rightarrow x'^\mu$ the gauge  field obviously transforms as
\begin{equation}\label{FF'}
F^{ij}_{\mu\nu}(x) \rightarrow
F'^{ij}_{\mu\nu}(x') = 
\frac{ \partial x^\lambda}{\partial x'^\mu} 
\frac{ \partial x^\rho}{\partial x'^\nu} 
F^{ij}_{\lambda\rho}(x).
\end{equation}
Moreover, a coordinate transformation does not affect the action.
From this one deduces that the matrix $H^{i\mu}(x)$ is transformed
by the coordinate transformation as
\begin{equation}\label{Hh}
H^{i\mu}(x)\rightarrow h^{i\mu}(x') = 
\frac{ \partial x'^\mu}{\partial x^\lambda}  H^{i\lambda}(x).
\end{equation}
We chose different notation for the transformed matrix
calling it $h^{i\mu}(x)$ in order to distinguish it
from  the unimodular matrix $H^{i\mu}(x)$. According to
(\ref{Hh}) the determinant of the transformed matrix
depends on the coordinate transformation 
\begin{equation}\label{deth}
\det [h^{i\mu} (x') ]
= \det \left[ \frac{ \partial x'^\mu}{\partial x^\lambda}\right].
\end{equation}
From Eqs.(\ref{FF'}),(\ref{Hh}) one deduces that in arbitrary
coordinates $x$ the action (\ref{HHF}) can be presented in the following
form
\begin{equation} \label{hhF}
\Delta S = -f \int h^{i\mu}(x)h^{j\nu}(x)F^{ij}_{\mu\nu}(x) \det h(x)
\,d^4x.
\end{equation}
It is convenient for further discussion
to introduce notation in which $h^{i}_\mu(x)$ is understood
as the matrix  
inverse to $h^{i\mu}(x)$,  i.e.
\begin{equation}\label{h-1} 
h^{i}_\mu(x)h^{j\mu}(x) = \delta_{ij}.
\end{equation}
The determinant in Eq.(\ref{hhF}) is defined as a determinant of
this inverse matrix.
Thus the factor $\det h(x)$ in  (\ref{hhF})
simply accounts for the variation of the
phase volume under the coordinate transformation
\begin{equation}\label{d^4}
\det h(x') d^4 x'
\equiv \det [ h^i_\mu(x')] d^4 x'=  \det 
\left[ \frac{ \partial x^\mu}{\partial x'^\nu} \right] d^4 x' = d^4x,
\end{equation}
Here $x$ and $x'$ are the orthogonal and 
arbitrary coordinates respectively.

Summarizing, it is shown in this Section that
(\ref{hhF}) gives the effective action which describes the 
interaction of a weak and smooth gauge field
$F^{ij}_{\mu\nu}(x)$ with polarized instantons. 

\section{The Riemann geometry and the Einstein equations}
\label{eineq}
Let us consider in this Section excitations above 
the polarized vacuum in the classical approximation. 
It is clear that these excitations 
should have very interesting properties 
because  a variation of the gauge field results in the contribution
to the action (\ref{hhF}) 
which is linear in the  field. This is in contrast to
the standard quadratic behavior of the Yang-Mills action (\ref{yma}).
We will assume that the fields considered
vary on the macroscopic distances, say $\sim 1$cm, and 
their magnitude can be roughly estimated as 
$|F^{ij}_{\mu\nu}| \sim 1/ {\rm cm^2}$.
We will see below that the constant $f$ which was
defined in Eqs.(\ref{M^AB}),(\ref{M=fM}) is  large,
$f \sim 1/r_{\rm P}^2$, where $r_{\rm P}$ is the Planck radius.
This shows that for weak fields the integrand in the 
Yang-Mills action (\ref{yma})  is suppressed compared to (\ref{hhF})
by a drastic factor 
\begin{equation}\label{frp}
|F^{ij}_{\mu\nu}|/f \sim (r_{\rm P}/1 {\rm cm})^2 = 10^{-64}.
\end{equation}
This estimate demonstrates that
our first priority is to take into account
the action (\ref{hhF}) which describes interaction
of the weak field with polarized instantons, neglecting the Yang-Mills
action (\ref{yma}). 
We will return to this point  in Section \ref{con}.

Thus we suppose that  properties of
low-energy excitations of the gauge field
can be described by the action (\ref{hhF}).
The fact that 
the field $F^{ij}_{\mu\nu}(x)$  varies on the macroscopic 
scale  makes it  weak and smooth on the microscopic level.
These properties of the field mean that it reveals trivial
topological structure on the microscopic level.
In contrast, the matrix $h^{i\mu}(x)$ describes
those degrees of freedom of the gauge field
which are associated with instantons and
therefore have highly nontrivial microscopic topological structure. 
Thus $F^{ij}_{\mu\nu}(x)$ and $h^{i\mu}(x)$ describe 
quite different topological structures.
Their different topology enables us  to consider 
them as two sets of independent variables, or modes. 
This makes the 
action (\ref{hhF})  a functional of these  variables
\[ \Delta S =\Delta S( \{A^{ij}_\mu(x)\},\{ h^{i\mu}(x)\}),\]
where  $A^{ij}_\mu(x)$ is the vector potential of the
external field $F^{ij}_{\mu\nu}(x)$.
We see that 
classical equations for the functional (\ref{hhF}) read
\begin{eqnarray} \label{dsda} 
\frac{\delta( \Delta S)}{ \delta A^{ij}_{\mu}(x)} &=& 0,
\\ \label{dsdh}
\frac{\delta( \Delta S)}{ \delta h^{i \mu}(x)} &=& 0.
\end{eqnarray}
From Eq.(\ref{hhF}) one finds that the first classical equation
(\ref{dsda})  results in
\begin{equation} \label{dA}
\nabla^{ik}_\mu [ \left(h^{k\mu}(x)h^{j\nu}(x)-
h^{k\nu}(x)h^{j\mu}(x)\right) \det h(x)] = 0.
\end{equation}
Here $\nabla^{ij}_\mu=\partial_\mu\delta_{ij}+
A^{ij}_\mu(x)$ is the covariant derivative
in the external gauge field.
The second classical equation which follows from
Eqs.(\ref{hhF}),(\ref{dsdh}) reads
\begin{equation} \label{eint}
h^{j\nu}(x)F^{ij}_{\mu\nu}(x) - \frac{1}{2}h^i_\mu(x)
h^{k\lambda}(x)h^{j\nu}(x)F^{kj}_{\lambda\nu}(x)  = 0.
\end{equation}
Here $h^i_\mu(x)$ is defined in Eq.(\ref{h-1}).
In order to present Eqs.(\ref{dA}),(\ref{eint})
in a more convenient form let us define three
quantities, $g_{\mu \nu}(x), 
\Gamma^{\lambda}_{\mu \nu}(x)$, and 
$R^{\lambda}_{\rho \mu \nu}(x)$:
\begin{eqnarray} \label{ghh}
g_{\mu \nu}(x) &=& h^{i}_{\mu}(x) h^{i}_{\nu}(x),
\\ \label{gam}
 \Gamma^{\lambda}_{\mu \nu}(x) &=& 
h^{i \lambda}(x)h^{j}_{\mu}(x)A^{ij}_
{\nu}(x) + h^{i \lambda}(x) \partial_{\nu} h^{i}_{\mu}(x),
\\ \label{RF}
R^{\lambda}_{\rho \mu \nu}(x) &=& h^{i \lambda}(x) 
h^{j}_{\rho}(x)F^{ij}_
{\mu \nu}(x).
\end{eqnarray}
Remember that space-time under consideration is basically 
flat. Therefore 
Eqs.(\ref{ghh}),(\ref{gam}),(\ref{RF}) just define 
the left-hand sides in terms of $A^{ij}_\mu(x)$ and $h^{i\mu}(x)$.
Remarkably, the classical Eqs.({\ref{dA}),(\ref{eint})
for the gauge field supply  these definitions with 
an interesting geometrical content.
In order to see this let us notice 
that after simple calculations
the first classical 
Eq.(\ref{dA}) may be presented in the following  form
\begin{equation} \label{Gg}
\Gamma_{\lambda\mu}^{\sigma}(x)=\frac{1}{2} 
g^{\sigma\tau}(x) \left[
\partial_\lambda g_{\tau \mu }(x)+
\partial_\mu g_{\lambda \tau }(x)-
\partial_\tau g_{\lambda \mu }(x)\right].
\end{equation}
Here the matrix $g^{\mu\nu}(x)$ is defined as
\begin{equation}\label{gin}
g^{\mu\nu}(x) = h^{i\mu}(x)h^{i\nu}(x),
\end{equation}
which according to (\ref{h-1}) makes it
 inverse to $g_{\mu\nu}(x)$
\begin{equation}\label{inve}
g^{\mu\lambda}(x)g_{\lambda\nu}(x) = \delta_{\mu\nu}.
\end{equation}
The form of Eq.(\ref{Gg}) is identical to the usual
relation  which expresses the Christoffel symbol
in terms of the Riemann metric for some Riemann geometry
\cite{LL}.
Moreover, using
({\ref{Gg}),(\ref{ghh}),(\ref{gam}), and (\ref{RF}) 
it is easy to verify that the quantity
$R^{\lambda}_{\rho\mu \nu}(x)$ can be presented in 
terms of $g_{\mu\nu}(x)$ as well
\begin{equation} \label{RG}
R_{\lambda\mu\nu}^{\lambda}(x)=
\partial_\mu \Gamma_{\lambda\nu}^{\lambda}-
\partial_\nu \Gamma_{\lambda \mu}^{\lambda}+
\Gamma_{\tau \mu}^{\sigma}\Gamma_{\lambda\nu}^{\tau}-
\Gamma_{\tau\nu}^{\sigma}\Gamma_{\lambda\mu}^{\tau}.
\end{equation}
%% 1
One recognizes in this 
relation the usual connection between the Riemann
tensor and the Riemann metric. Furthermore,
it follows from (\ref{Gg}) that 
$g_{\mu\nu}(x), \Gamma^\lambda_{\mu\nu}(x)$ and 
$R_{\lambda\mu\nu}^{\sigma}(x)$ possess the symmetry properties
which are usual in the Riemann geometry
\begin{eqnarray}\label{gsy}
g_{\mu\nu}(x) &=& g_{\nu\mu}(x), \\ \label{Gsy}
\Gamma^\lambda_{\mu\nu}(x) &=& \Gamma^\lambda_{\nu\mu}(x), \\ \label{Rsy}
R_{\sigma\rho\mu\nu}(x)&=& R_{\sigma\rho\nu\mu}(x)=R_{\rho\sigma\mu\nu}(x)=
R_{\mu\nu\sigma\rho}(x),
\end{eqnarray}
where 
\[R_{\sigma\rho\mu\nu}(x)=
g_{\sigma\lambda}(x)R^\lambda_{\rho\mu\nu}(x).\]
Eqs.(\ref{Gg}),(\ref{RG}), (\ref{gsy}),(\ref{Gsy}) and (\ref{Rsy})
show that $g_{\mu\nu}(x)$ can be considered as 
a metric for some Riemann geometry 
with the Christoffel symbol $\Gamma^\lambda_{\mu\nu}(x)$
and the Riemann tensor $R^\lambda_{\rho\mu\nu}(x)$.

Consider now the second classical equation (\ref{eint}).
With the help of Eqs.(\ref{ghh}),(\ref{RF}) it is easy to verify that
it can be presented
in the following form
\begin{equation} \label{einnom}
R_{\mu \nu} - \frac{1}{2} g_{\mu \nu} R = 0.
\end{equation}
Here 
\begin{eqnarray}\label{ric}
&&R_{\mu\nu}(x) = R^\lambda_{\mu\lambda\nu}(x), \\ \label{kriv}
&&R(x) = g^{\mu\nu}(x) R_{\mu\nu}(x).
\end{eqnarray}
These definitions show that
$R_{\mu\nu}(x)$ and $R(x)$ are the
Ricci tensor and the curvature of the Riemann geometry based on the
metric $g_{\mu\nu}(x)$. 
Hence, Eq.(\ref{einnom}) proves be identical to the
Einstein equations of general relativity in the absence of matter.

We come to the interesting result.
Remember that the matrix  $h^{i\mu}(x)$ describes the orientation
of (anti)instantons in the considered nontrivial vacuum, 
while $A^{ij}_\mu(x)$ is the potential of the weak gauge field.
Both these quantities describe  properties of the gauge field.
We  have demonstrated above that the 
first classical condition (\ref{dsda})
for the gauge field
can be expressed as a statement that particular combinations
(\ref{ghh}),(\ref{gam}),(\ref{RF}) 
of $h^{i\mu}(x)$ and $A^{ij}_\mu(x)$ 
are identical to the Riemann metric,
the Christoffel symbol, and the Riemann tensor for some Riemann space.
The second classical equation  (\ref{dsdh})
proves be identical to the Einstein equations for this Riemann metric.

The Einstein  equations imply in 
particular that there appear excitations called
gravitational waves. 
They possess zero mass and spin-2,
exactly what we have been looking for.
In the considered above picture these excitations arise 
from variables which describe the particular degrees
of freedom of the gauge field.
Remember that these variables  can be considered as the two  modes.  
One of them  describes a
weak  topologically trivial gauge field. 
The other one describes polarization of instantons.
The strong interaction of these two modes 
(\ref{hhF}) mixes them and 
the resulting spin-2 excitation should
be considered as a coherent propagation of 
the two interacting modes.

It is instructive to consider 
the action (\ref{hhF}) when the first classical  Eq.(\ref{dsda}) 
is valid. It is clear from (\ref{ghh}),(\ref{RF}) 
that the integrand of the action  (\ref{hhF})  
proves be
proportional to the integrand of the usual
Hilbert action of general relativity \cite{LL}  which 
has the following form in Euclidean formulation
\begin{equation}\label{hil}
S_{H}= -\frac{1}{16\pi k} \int R(x) [\det g(x)]^{1/2}\,d^4x.
\end{equation}
One can consider the two actions 
(\ref{hhF}) and (\ref{hil}) as same quantity
if the Newton gravitational constant $k$ is connected with
the constant $f$ which characterize the density of the 
polarized condensate of instantons
\begin{equation} \label{newt}
k = \frac{1}{16 \pi f}.
\end{equation}
This relation demonstrates that $f=2/r_{\rm P}^2$, thus supporting
estimation (\ref{frp}). Remember that the constant 
$f$ introduced in (\ref{M^AB}),(\ref{M=fM})
depends on the typical
radiuses and separations of the polarized instantons.
Relation (\ref{newt}) shows that we should assume that
these radiuses and separations are comparable with the Planck
radius.

The consideration given verifies that
the long-range excitations of the gauge field 
proves to  be the massless spin-2 particles.
Their classical propagation is described by
the Einstein equation of general relativity. The corresponding
effective action turns out to be identical to the  
Hilbert action of general relativity.
These facts altogether permit one to identify the found excitations 
with gravitational waves.
This indicates that
gravity arises in the framework of the gauge theory.
It is very important that  the dynamics of general relativity,
its action and equations of motion,
originate directly from the dynamics of the gauge field.
All these results follow directly from 
assumptions  (\ref{M^AB})-(\ref{Sig}) which 
were interpreted above as the $SO(3,3)$ polarization of instantons.

The picture developed above 
remains valid until the gauge
field varies smoothly on the radius of polarized instantons.
Under this condition the action (\ref{hhF}) remains valid.
As was mentioned,
radiuses of polarized instantons
are comparable with the Planck radius.
It makes it necessary for 
gravitational waves to have the wavelength
larger then the Planck radius.
For shorter wavelengths the term (\ref{hhF}) 
does not manifest itself in the action.
This means  that
the interaction of short wavelength gauge field
with the polarized instantons is suppressed. 
Therefore in this high-energy region 
the  gauge field is described by the conventional
Yang-Mills action and reveals its usual properties.
In particular its excitations are 
spin-1 gauge bosons, while gravitons do not exist.
Thus gravity manifests itself only 
for energies below the Planck energy, while for
higher energies it does not exist.

Remember that $h^{i\mu}(x)$ was introduced in Section \ref{eineq}
as an order parameter which describes the polarization
of instantons. 
It is interesting that
the Riemann structure defined 
in (\ref{ghh}),(\ref{gam}),(\ref{RF}) 
makes it possible another physical interpretation
of this quantity.
Namely, $h^{i\mu}(x)$ can be  identified 
with the vierbein, the quantity which generally
speaking is  well known in general relativity
\cite{LL}.  
The way of deriving  the Einstein equation
(\ref{einnom}) from the two classical equations
is similar to the Palatini method, see Ref.\cite{LL},
formulated with the help of the vierbein formalism.
Our consideration reveals however an important subtlety.
In the usual vierbein formulation  the physical 
nature of the space where the index
$i$ of the matrix $h^{i\mu}(x)$ belongs   
does not play a substantial role. This index can be
considered  purely as a label \cite{LL}.
In contrast, in our approach 
this index belongs to the isotopic space  
which plays the most important central role.
In this space gauge transformations of the
considered $SO(4)$  gauge theory are defined.
There is however a price for this
new physical interpretation of the index $i$.
It is important for us to rely on the Euclidean formulation.
An attempt to use the Minkowsky formulation 
as a starting point 
meets a difficulty.
Really, in Euclidean formulation a connection between 
the metric and the vierbein can be presented as 
\[ g^{\mu\nu}(x)h^i_\mu(x)h^j_\nu(x) = \delta_{ij},\] 
where Eqs.(\ref{ghh}),(\ref{inve}) were used. 
The delta-symbol in the right hand side
here simply shows that the isotopic space is
the 4D Euclidean space. In contrast, if one  would attempt to 
develop an approach used above
starting from the Minkowsky coordinate space then 
connection between the metric
and the vierbein would look  \cite{LL}
\[ g^{\mu\nu}(x)h^i_\mu(x)h^j_\nu(x) = g^{(0)ij} \]
where $ g^{(0)} = {\rm diag}(1,1,1,-1)$ is
the Minkowsky  metric. Thus 
the space which in our
approach should be considered as an isotopic space 
for some gauge theory acquires
the structure of the $3+1$ Minkowsky space.
Gauge  transformations in this space look 
as  $SO(3,1)$ transformations 
belonging to the noncompact Lorentz group. 
As  is well known,  the gauge theory for noncompact groups 
is not unitary  and therefore is
poorly defined \cite{gel}.
Thus the gauge formulation discussed above becomes
questionable  if the Minkowsky space would be used
 a starting point.

We conclude that 
an attempt to continue the vierbein into 
real space-time presents a problem.
Fortunately, one can avoid it. One can use
first Euclidean formulation, deriving
the Einstein equations (\ref{einnom}) as was discussed above. 
These equations are
formulated entirely in terms of  metric,
neither the vierbein nor any isotopic index
manifest themselves in these equations explicitly.
This fact permits one to fulfill  continuation 
of the final Einstein equations into real space-time
by continuation of the metric without 
reference to  the vierbein formalism.

In summary, it is shown in this Section that
the Einstein equations of general relativity
arise from of the $SO(4)$ gauge theory.

\section{Discussion of results}
\label{res}

The main results of this paper is a set of conditions
which are to be fulfilled in order to derive
the effect of gravity by means of conventional gauge theory.
Let us formulate  these results presenting them as comments to
the list of questions
posed in the Introduction.

1.It is shown above that
one can hope to deduce the effect of gravity from
the  gauge theory if the gauge group is $SO(4)$.

2.The necessary vacuum is shown to 
include the polarized instantons.
There is a simple physical reason explaining  a necessity for
this vacuum state.
Compare the Yang-Mills
action (\ref{yma}) with the Hilbert action (\ref{hil}). The striking
difference between them is the power of the field
in their integrand.
The Yang-Mills action is quadratic in the gauge field strength.
In contrast the Hilbert action (\ref{hil})  
can be considered as a quantity which
is linear in the Riemann tensor due to an 
obvious  relation  $R = g^{\mu\nu}R^\lambda_{\mu\lambda\nu}$.
At the same time there is the long-known resemblance
between the basic properties of the gauge field strength and
the Riemann tensor, see for example Ref.\cite{fad}. 
This paper proposes to make this resemblance
identity in a sense explained in (\ref{RF}). 
This proposal can only be accomplish
if some phenomenon in gauge theory produces
a term in the action  which is linear in the field strength.
Eq.(\ref{cdg}) shows that an instanton 
provides just the necessary effect. 
Moreover, an  ensemble of polarized instantons 
results in the action (\ref{hhF}) which proves to be identical
to the Hilbert action. This makes polarized instantons be so 
special for gravity.

3.It is argued above that the role of the order parameter
for the phase with polarized instantons 
should is played by mean orientations of instantons and
antiinstantons.
The orientation for an (anti)instanton is described by 
$3\times 3$ matrixes. 
Therefore in the  $SO(4)$
gauge theory  there can arise four  $3\times 3$  matrixes
describing mean orientations of all available 
instantons and antiinstantons. 
The main assumption of this paper is  that
these four matrixes can be described by one
$6\times 6$ matrix which belongs to $SO(3,3)$.
It is important to mention that one cannot afford something
less than that, the  order parameter 
in the picture suggested has to be an
$SO(3,3)$ matrix. To see this one can reverse arguments of
Section \ref{eineq}.
The gravitational field is known to be described by
the vierbein which can be looked at as a $SL(4)$ matrix.
Therefore to describe gravity by means of gauge theory
one needs to express  the vierbein, the local $SL(4)$ matrix
in terms of some variables of gauge theory. 
Eq.(\ref{hom}) provides such a
possibility, provided there exists the $SO(3,3)$ order parameter.
This consideration shows in particular that the 
gauge theory should be based on the $SO(4)$ group (or bigger one)
which possesses sufficient number of different
(anti)instantons in order to
develop  the necessary $SO(3,3)$ order parameter.
 
4. Excitations above the considered vacuum
state are found  to be identical to gravitational waves.
These excitations are described 
in terms  of two modes of the gauge field.
One of them is a weak
topologically trivial gauge field $A^{ij}_\mu(x)$. The other mode
describes orientations of instantons $h^{i\mu}(x)$.
Eq.(\ref{hhF}) shows that there is the strong interaction
between these modes which  results in their  mixing.
Thus gravitational waves should be considered as a 
coherent propagation of the 
two interacting modes describing two different
degrees of freedom of the gauge field.

5.We verified that  low energy degrees of freedom
of the gauge field are described by the 
variables of the Riemann geometry if the
polarized vacuum exists. The low-energy action (\ref{hhF}) 
is found to be identical to the Hilbert action of general relativity.
This makes the classical equations be identical to
the Einstein equations of general relativity.
Eq.(\ref{newt}) shows that the relevant instantons
should have radiuses comparable with the Planck radius.
This fact restricts the energies from above.
For energies well below the Planck radius
the long-wave-length action (\ref{hhF}) is valid, 
resulting in existence of gravitational waves.
For energies  higher than the Planck limit 
Eq.(\ref{hhF}) becomes not applicable,
the weak gauge field  interacts
with instantons no more
and one should describe the gauge field
by the usual Yang-Mills action.
This shows that gravitational waves exist
only for low energies, while high energy behavior of
the gauge field should be  described by the gauge bosons.

6.We have not considered quantum corrections above.
From experience in QCD one could anticipate that they
might be dangerous for the scenario considered.
Even if the gauge constant is small for
some short distances, quantum corrections
in the QCD vacuum are known to
make it rise for larger distances 
\cite{kh,gross,poli} eventually resulting in confinement.  
One should keep in mind, however, that in the
considered polarized vacuum the role
of quantum corrections differ qualitatively
from the QCD vacuum.
Remember that the rise of the gauge coupling
constant in the QCD vacuum 
is related to those quantum fluctuations 
whose wavelength is the longer
the larger the distances considered.
The point is that the long range quantum fluctuations
of the gauge field in the considered vacuum 
are strongly affected by
polarized instantons. This makes the quantum corrections 
different from the ones in the QCD vacuum.
This is an important issue and it is worth to
repeat it in terms of the basic excitations.
The perturbation series in the QCD vacuum
is known to become non-reliable for large distances
due to large contribution  from  virtual  long-wavelength gauge
bosons. The divergence of the perturbation theory is usually 
considered as an indication that
for large  distances  the quantum corrections
become very important. 
In the polarized vacuum this picture does not work because, 
as we verified above, 
low-energy spin-1 gauge bosons simply do not exist in 
this vacuum.
There are spin-2 gravitational waves instead of them.
Definitely this argument deserves to be considered
in much more detail, which puts it outside the 
main stream of this paper.

7. Eq.(\ref{so33}) postulates the 
$SO(3,3)$ symmetry for the order parameter.
This proves to be  a very demanding assumption.
To illustrate this statement consider
the simplest possible case of zero gravitational field.
Notice  that choosing the Galilean
coordinates one can always eliminate the
gravitational field locally. This makes
the case of the zero field be of interest
for nonzero gravitational fields as well.
In the absence of gravitational field the vierbein
can be chosen to be a constant $SO(4)$ matrix
$h^{i\mu}(x) = H^{i\mu} \in SO(4)$.
Eq.(\ref{hom}) shows that the matrix $M^{AB}(x)$ 
in this case is a constant $SO(3)\times SO(3)$ matrix
\begin{equation}\label{so3so3}
M = 
\left(  \begin{array}{cc} C  & \hat 0   \\ 
\hat 0  &  \bar C \end{array} \right),
\end{equation}
where $C, \bar C \in SO(3)$.  This means according to (\ref{T})
that the antiinstantons in the $su(2)\eta$
subalgebra must be polarized, their polarization
is described by the orthogonal matrix $C$.
Similarly, the instantons in the $su(2)\bar\eta$
subalgebra are also polarized, their polarization
is described by the orthogonal matrix $\bar C$.
In contrast, the instantons in the $su(2)\eta$ and 
antiinstantons in the $su(2)\bar\eta $ subalgebras
are not polarized. 
Thus the polarization of (anti)instantons is to be not
trivial even in the absence of the gravitational field.

To meet the requirement (\ref{so3so3})
one should develop  a $SU(2)$ gauge theory model in which
instantons  are polarized, while
antiinstantons remain  not polarized. 
The candidate for such a model was proposed in 
Refs.\cite{det,PRdet}.
In this model there arises interaction between instantons
which makes their identical orientation more  probable.
This interaction originates from single-fermion loop
correction when interaction of fermions with the given
instantons as well as with 
particular scalar condensates is taken into account.
There is no interaction of this type between antiinstantons.
The interaction between instantons
provides  a possibility for a phase transition
into the state with polarized instantons.
An existence of the polarized  phase 
has been verified in \cite{PRdet} in the mean field approximation.
The order parameter for polarized instantons
has the $SO(3)$ symmetry as is necessary to satisfy
Eq.(\ref{so3so3}).

Thus the mentioned model gives a hope that 
the phase described by 
the simplest $SO(3)\times SO(3)$ order parameter (\ref{so3so3}) 
can exist.
To move further on one should overcome two 
limitations of the model.
Firstly, the way is to be found to make the ensemble
of instantons and antiinstantons in the $SO(4)$ gauge
theory to develop the condensate governed by the
more sophisticated $SO(3,3)$ order parameter.
Secondly, the model of \cite{det,PRdet}
relies upon existence of the scalar condensates transformed
by the gauge group. 
These condensates make the gauge field
acquire a mass through the Higgs mechanism.
Therefore there is a danger that the mass of the gauge
field would result in creation 
of the  mass for gravitational waves.
One should  find a way to avoid this undesirable property.

Thus, either development of the model \cite{det,PRdet}
or possibly  a new more sophisticated model is necessary.
The present paper clearly sets the requirements for such model.

8.The only information available on  possible phase transitions
into the state with polarized instantons comes from the
mentioned above model  \cite{det,PRdet} in which
it was found that in the mean field approximation
there exists the
first order phase transition into the polarized  state.

Our last comment is on the dilute gas
approximation for instantons.
It is shown in  Appendix \ref{A} that 
even  closely located instantons interact
with the external field as individual objects. 
This fact may be considered as an
indication  that the picture developed above
for the gas of instantons may be valid for
the instanton liquid as well.

\section{Conclusion}
\label{con}
The conventional approach to  general relativity postulates
that the theory is to be written in geometrical terms. 
Postulating then the Hilbert action
one derives the Einstein equations. 
In this paper we discuss a possibility for a novel approach. 
Space-time is supposed to be basically flat.
In this space-time the $SO(4)$ gauge theory is formulated.
Assuming that in this theory there develops the particular phase,
called polarized instantons we find that  for low energies 
the gauge field can be described in terms of the Riemann geometry.
There are two low-energy modes of the gauge field which play the
important role.
One of them accounts for topologically
trivial sector of the theory,  the other one 
describes the polarization of instantons.
These two modes  strongly
interact.  Their interaction originates
from the conventional Yang-Mills action, but 
for weak fields it can be presented in the form of the Hilbert
action from which  one derives the Einstein equations
describing gravitational waves. Thus the gravitational
wave arises as a mixing of the two modes 
of the gauge field.

\ack
I am thankful to
R.Arnowitt,
C.Hamer, 
E.Shuryak
for  critical comments.
My participation in the
Workshop on Symmetries in the Strong Interaction, Adelaide 1997,
where part of this work was fulfilled, was helpful.
The support of the Australian Research Council is appreciated.

\appendix

\section{Interaction of instantons with an external field}
\label{A}

Let us show that instantons interact with an 
external field as  individual objects. It means  that
the action describing 
the   interaction of several instantons with
a weak slowly varying external gauge field
is given by a sum of terms 
of Eq.(\ref{cdg})  type
describing
interactions of different instantons with the field.
This statement is trivial when the dilute gas
approximation is valid. Surprisingly it remains valid
for any configuration of  instantons, even if
they are overlapping.

To describe the closely situated instantons one needs
to consider the  $k$-instanton general  solution \cite{ADHM},
for the review see  \cite{pras}.
It is sufficient for our purposes to suppose that
all instantons are in some $su(2)$ subalgebra.
Let us introduce a quaternion as $q=q_\mu \tau^+_\mu$, where
 $q_\mu, \mu=1, \cdots,4$ is an arbitrary vector. In this 
notation 
$q^+=q_\mu \tau^-_\mu$, where $\tau^\pm_\mu = 
(\pm i \vec \tau,1)$.
Consider the $(k+1)\times k$ matrix $M_{s,t}(x),
s=1,\cdots,k+1,~t=1,\cdots,k$ which has quaternionic matrix 
elements.  
$M(x)$ is a linear function of the quaternion of coordinates 
$x=x_\mu \tau^+_\mu$
%?\begin{equation} \label{M(x)}
\begin{displaymath}
M(x)=B-Cx,
\end{displaymath}
where $B$ and $C$ are $x$-independent $(k+1)\times k$ 
quaternionic matrixes.
They must be chosen so that the condition
\begin{equation}\label{Rx}
M^+(x)M(x)=R(x)
\end{equation}
be fulfilled for any $x$. 
Here $R(x)$ is a non-degenerate $k\times k$ real matrix.
The matrix $C$ can be chosen to be
\begin{equation}\label{Cmatr}
C_{1t}=0,~ ~C_{1+s,t}=\delta_{st}, ~s,t=1,\cdots,k.
 \end{equation}
(The matrix $M(x)=M_{s,t}(x)$ which plays the central role
in the ADHM construction should not be confused with the
matrix $M(x) = M^{AB}(x)$ defined in 
(\ref{M=fM}) to describe polarization of instantons.)
Having $M(x)$ one can find the $k+1$ quaternionic vector 
$N(x)$ which
satisfies equations
\begin{eqnarray}\label{MN}
M^+(x)N(x)&=&0,\\ \label{NN}
N^+(x)N(x)&=&1.
\end{eqnarray}
Then the vector-potential defined in the  quaternion 
representation as
\begin{equation} \label{Ains}
A_\mu(x)=\frac{1}{2i} A_\mu^a(x)\tau^a=N^+(x)\partial_\mu N(x)
\end{equation}
results in  the general self-dual gauge field for $k$ 
instantons
\begin{eqnarray}\label{fmn}
F_{\mu \nu}(x)&=&\frac{1}{2i} 
F^a_{\mu\nu}(x)\tau^a \\ \nonumber
&=&2i N^+(x)C\eta^{a\mu\nu} \tau^a R^{-1}(x)C^+N(x).
\end{eqnarray}
To simplify notation in the following consideration
consider the case of two instantons
when the matrixes  $M(x),C$ 
in the ADHM solution have an explicit simple form
\begin{equation}\label{matM}
M(x) = 
\left(  \begin{array}{cc}
q_1 & q_2\\
y_1-x & b\\
b&y_2-x
\end {array}\right),~~~~
C = 
\left(  \begin{array}{cc}
0 & 0\\
1 & 0\\
0 & 1
\end{array}\right),
\end{equation}
where $q_i,~i=1,2$ are the quaternions describing the radiuses 
$\rho_i= |q_i|$
and orientations of the two instantons.
The orientations of instantons
may be considered as $SU(2)$ matrixes $n_i=q_i/|q_i|$.
An equivalent definition
of the orientation may be given in terms of a
matrix $D^{ab}\in SO(3)$, see Eq.(\ref{cdg}).
The relation 
\begin{equation} \label{CU}
\tau^a D^{ab}_i= n_i\tau^b n^+_i~
\end{equation}
defines these $SO(3)$ orientation matrixes for the two instantons
in terms of $q_i$.
The positions of the instantons  are given by $y_i$, and
$b$ is 
\begin{equation}\label{b}
b=\frac{y_{12}}{2 |y_{12 }|^2}
\left(q_2^+q_1-q_1^+q_2\right),~~
y_{12}=y_1-y_2.
\end{equation}
In the following consideration we will need
the instanton field in the region $x$
\begin{equation} \label{xy}
|x-(y_1+y_2)/2|
\gg |y_{12}|,\rho_1,\rho_2.
\end{equation}
Eq.(\ref{MN}) for two instantons reads
\begin{eqnarray}\label{N1}
(x-y_1)^+N_2(x) = q_1^+N_1(x)+b^+N_3(x), \\ \label{N2}
(x-y_2)^+N_3(x) = q_2^+N_1(x)+b^+N_2(x).
\end{eqnarray}
The approximate solution of this equation in the 
region (\ref{xy}) is
\begin{eqnarray}\label{Nas1}
N_2(x) \simeq \frac{x-y_1}{|x-y_1|^2} q^+_1  N_1(x),
\\ \label{Nas2}
N_3(x) \simeq \frac{x-y_2}{|x-y_2|^2}q^+_2 N_1(x),
\end{eqnarray}
where the terms $\sim 1/|x-y_i|^2$ are omitted in the lhs's.
From these relations and the normalization condition 
(\ref{NN})
one finds $|N_1(x)|  \simeq 1$. A choice
\begin{equation} \label{N11}
N_1(x)=1,
\end{equation}
fixes the singular gauge.
The matrix $R(x)$ for two instantons in the
region (\ref{xy}) is simplified to be
\begin{equation}\label{R(x)}
R(x) \simeq \left( \begin{array}{cc}
|x-y_1|^2 & 0 \\
0         & |x-y_2|^2
\end{array} \right).
\end{equation}
Substituting 
(\ref{Nas1}),(\ref{Nas2}),(\ref{N11}),(\ref{R(x)}) 
in Eqs.(\ref{Ains}),(\ref{fmn}) 
one finds that the potential and the
field created by the two instantons in the region
(\ref{xy})  is equal to the sum of 
potentials and fields created by individual instantons
\begin{eqnarray}\label{sum}
A^a_{\mu\nu}(x)= \sum_{i=1,2}A^a_{\mu\nu,i}(x),\\ \label{sumF}
F^a_{\mu\nu}(x)=\sum_{i=1,2}F^a_{\mu\nu,i}(x).
\end{eqnarray}
Here $A^a_{\mu\nu,i}(x),F^a_{\mu\nu,i}(x)$ are given
by the known expressions which in the region (\ref{xy})
read
\begin{eqnarray}\label{indA}
A^a_{\mu\nu,i}(x)&=&2 \bar \eta^{b\mu\nu} D^{ba}_i
\frac{(x-y_i)_\nu }{|x-y_i|^4}\rho_i^2,\\ \label{andF}
F^a_{\mu\nu,i}(x)&=&-4
M_{\mu\mu',i} M_{\nu\nu',i}
 \bar \eta^{b\mu'\nu'} D^{ba}_i \frac{1}{|x-y_i|^4}
\rho_i^2.
\end{eqnarray}
Here $M_{\mu\nu,i} = \delta_{\mu\nu}- 
2(x-y_i)_\mu (x-y_i)_\nu/|x-y_i|^2$. 
Thus in the region (\ref{xy}) the instantons contribute
independently to the gauge potential and the gauge field
even for small separations of instantons.

Consider now the interaction of the
external gauge field $[F_{\mu\nu}^a]^{\rm ext}$
with two instantons generalizing the approach of \cite{CDG}
originally proposed for a single instanton.
The field is supposed to be  weak  enough
$|\,[F_{\mu\nu}^a]^{\rm ext}\,| \ll 1/(g^2\rho_1^2),
1/(g^2\rho_2^2)$.
It is also supposed to vary slowly, 
$r | \partial_\lambda [F_{\mu\nu}^a]^{\rm ext}|
\ll |\,[F_{\mu\nu}^a]^{\rm ext}\,|$, where $ r ={\rm max}
(\rho_1,\rho_2,|y_{12}|)$.
Consider the sphere of  radius $R \gg r$ 
centered at $x_0=(y_1+y_2)/2$
 surrounding the
instantons. Let us choose the radius of the sphere $R$ 
to be so small that the variation of the external
field inside the sphere is negligible. At the same
time the radius $R$ is to  be large enough to
make the external field outside the sphere larger
than the field created by the instantons.
Let us use the Landau gauge for the potential
of the external 
field 
\begin{equation} \label{Lg}
[A_\mu^a(x)]^{\rm ext}=-\frac{1}{2}(x-x_0)_\nu
[F^a_{\mu\nu}(x_0)]^{\rm ext}.
\end{equation}
Consider the action of the gauge field 
\begin{equation} \label{agf}
S= \frac{1}{4g^2}\int F^a_{\mu\nu}(x)F^a_{\mu\nu}(x)\,d^4x.
\end{equation}
The instantons give large contribution to the action
$S^{\rm ins} = 2\times (8 \pi^2/g^2)$ which comes
mainly from the inner-sphere region of integration
in (\ref{agf}). Let us
find now the contribution which  the inner-sphere region
gives to the action (\ref{agf}) where the external field
can be considered as a perturbation. From the definition
%?\begin{equation} \label{SI}
\begin{displaymath}
S_I(\{F\}) =\frac{1}{4g^2}\int_{|x-x_0|<R}
F^a_{\mu\nu}(x)F^a_{\mu\nu}(x)\,d^4x
\end{displaymath}
one finds
\begin{equation} \label{delSI}
\delta S_I = \frac{1}{g^2}\int_{|x-x_0|<R}
\nabla_\mu \left([A^a_\nu(x)]^{\rm ext}\right)
[F^a_{\mu\nu}(x)]^{\rm ins}\,d^4x. 
\end{equation}
Here $\delta S_I= S_I(\{F^{\rm ins}+F^{\rm ext}\})-
S_I(\{F^{\rm ins}\})$.
Integrating by parts and using the classical equation
$\nabla_\mu [F^a_{\mu\nu}(x)]^{\rm ins}=0$
for the field of instantons one finds from (\ref{delSI})
\begin{equation} \label{sphere}
\delta S_I = \frac{1}{g^2}\int
[A^a_\nu(x)]^{\rm ext} [F^a_{\mu\nu}(x)]^{\rm ins}
\,dS_\mu,
\end{equation}
where $\int dS_\mu$ denotes integration over the sphere.
Substituting  (\ref{sumF}),(\ref{andF}),(\ref{Lg}) 
in Eq.(\ref{sphere}) one finds
the contribution 
of the inner-sphere region to the action
\begin{equation} \label{SIsum}
\delta S_I=\frac{\pi^2}{g^2}\sum_{i=i,2}\rho_i^2
\bar \eta^{a\mu\nu}
D^{ab}_i F^b_{\mu\nu}(x_i).
\end{equation}
One can consider similarly the contribution $\delta S_E$
of the outer-sphere region where the field of instantons
plays the role of a perturbation. The integration
again is reduced to integration over the sphere
and the final result is identical to (\ref{SIsum}),
$\delta S_E=\delta S_I$.
Deriving it one should neglect
the integral over 4-dimensional outer-sphere region
\begin{equation} \label{outi}
\int_{|x-x_0|>R} 
[A^a_\nu(x)]^{\rm ins}\nabla_\mu [F^a_{\mu\nu}(x)]^{\rm
ext}\,d^4 x \rightarrow 0.
\end{equation}
There are two possible reasons for this. If the external
field is strong in some region,
then it is supposed to satisfy there
the  classical equation $\nabla_\mu
[F^a_{\mu\nu}(x)]^{\rm ext}=0$.
The weak field is supposed to
decrease smoothly at infinity. If $l$ is a large length
parameter governing this field, 
$|\,[F^a_{\mu\nu}]^{\rm ext}\,|
\sim 1/l^2,~|\,\nabla_\rho [F^a_{\mu\nu}(x)]^{\rm ext}\,|
\sim 1/l^3$, then 
the  integral in (\ref{outi}) is estimated
as $\int_{x\sim a} d^4x 
\sim (\rho_1^2+\rho_2^2)/l^2 \ll 1$.
Note that  the decrease
of the external field should be slower than
the decrease of the instantons field to ensure that
everywhere in the outer-sphere region the external field
exceeds the instanton field.

Summing contributions of both inner-region and outer-region
one finds
\begin{equation} \label{sevins}
\Delta S=\frac{2\pi^2}{g^2}
\sum_{i}\rho_i^2\bar \eta^{a\mu\nu}
D^{ab}_i F^b_{\mu\nu}(x_i).
\end{equation}
The external field considered above was supposed not
to vary dramatically
on the distance of instanton separation $|y_{12}|$.
This condition was assumed in order to take into 
consideration the case of small separations where the obtained
result is  surprising. If the separation
is larger then radiuses, 
than two instantons may certainly be considered
as  individual objects and (\ref{cdg})
may be applied to every instanton separately. In this case
there is no need to suppose that the external
field remains constant,
it may vary from one instanton to the other. 
This fact permits one to chose
the argument of the external field in (\ref{sevins}) 
as $x_i$.
Eq.(\ref{sevins}) is verified above for the case of  
two instantons  mainly to simplify notation.
A similar consideration shows that
it remains valid for any number of instantons
provided the external field
is weaker than the instanton field in the vicinity 
of each instanton and its variation on their radiuses is negligible.
The instantons separations may be arbitrary.

Remember that an instanton
provides a minimum for  the classical action 
for small variations of the field.
The action (\ref{cdg}) reveals a nontrivial behavior,
it is linear in the field.  
This property is explained by the fact that
the external field is stronger than
the field of instantons in the outer-sphere region.
Therefore in this region the perturbation theory breaks down.
As a result there appears the 
surface-term (\ref{sphere}) in the 
action which describes the
integration over a 3-dimensional surface
separating the region where the external field is strong.
It is  this term that gives rise to the action (\ref{sevins}).
In the approach developed in  \cite{VZNS} the surface-term   
does not manifest itself explicitly. There exists
however the field of free gauge 
bosons created by the instanton. 
Their radiation field  exceeds the field of the 
instanton at infinity.

In conclusion, we verified that
instantons interact with the external
field independently even if
their separations are small. Notice that this
result  demonstrates 
very clearly that the number of
parameters in the $k$-instanton ADHM-construction 
is equal to $8k-3$ and therefore this construction
describes the most general configuration of instantons. 
In Ref.\cite{CWS}  this statement was proved 
using more sophisticated arguments.

\section{ ${\bf SO(3,3) \protect\cong  SL(4) }$}
\label{B}

Let us describe properties of  the $ SO(3,3)$ group,
in particular a  relation  which connects the $ SO(3,3)$  and $SL(4)$  
groups.
An  existence of this relation
is known in mathematical literature, see for example Ref.\cite{gil},
where it is mentioned. The goal of this Section 
is to show that  this relation has the form
defined by  (\ref{hom}).

Notice first that any matrix 
${M}^{AB},~A,B = 1,2, \cdots,6,~{M}\in SO(3,3)$
may be  presented in the following form
\begin{equation} \label{ccdd}
{M} = 
\left(  \begin{array}{cc} C & \hat 0 \\ \hat 0 & \bar C 
\end{array} \right)
\left(  \begin{array}{cc} \sigma U & \bar \sigma V\\ 
                          \bar \sigma V & \sigma U \end{array}
 \right)
\left(  \begin{array}{cc} 
D^T &\hat 0 \\\hat 0 & \bar D^T \end{array} \! \right).
\end{equation}
Here $C,\bar C,D,\bar D \in SO(3)$ and $U,V$ are diagonal 
$3\times 3$ 
non-negative matrixes
%?\begin{equation} \label{uv}
\begin{displaymath}
U_{ij} = U_i \delta_{ij},~~
V_{ij}=V_i \delta_{ij},~~U_i,V_i \ge 0,
\end{displaymath}
satisfying condition
%?\begin{equation} \label{uivi}
\begin{displaymath}
U^2-V^2 = \hat 1.
\end{displaymath}
$\sigma,\bar\sigma$ are the numbers: $\sigma =\pm 1,
\bar \sigma \pm 1$. Numbers with a hat are used to represent 
$3\times 3$
 matrixes.
The 15 parameters governing a matrix belonging to
$SO(3,3)$ are: 3 parameters for every one
of the four $SO(3)$ matrixes in (\ref{ccdd}) plus
3 parameters governing the matrixes $U,V$.

It is clear that if $\sigma=1$ in (\ref{ccdd})
then $M$ may be transformed into the unity $6\times 6$ 
matrix ${\bf 1}$  by the continuum transformation, 
 i.e. there exists the continuum function $M(s),~
M(s)\in SO(3,3),~s \in [0,1],~M(0)=M,~M(1)={\bf 1}$.
Therefore $\sigma =1$ implies $M\in SO_+(3,3)$. If
$\sigma =-1$, then this transformation is impossible,
this case is denoted as $M \in SO_-(3,3)$. 
The proof of Eq.(\ref{ccdd}) requires 
simple conventional algebraic transformations 
based on (\ref{MSM}).

Consider now
a matrix $H^{ij}\in SL(4)$ and construct from it the 
quantity $H^{im}H^{jn}-H^{in}H^{jm}$. 
The later one is obviously antisymmetric
in $i,j$ as well as in $m,n$. Therefore it
can be expanded in a series of the
't Hooft symbols which provide 
the full basis for  antisymmetric matrixes.
It is convenient to present this expansion   with the help of  
$\eta^{Aij},A=1,2,\cdots ,6$,  
symbols defined in (\ref{etaA}).
Then the expansion reads
\begin{equation} \label{hom1}
H^{im}H^{jn}-H^{in}H^{jm} =\frac{1}{2}{M}^{AB}\eta^{Amn}
\eta^{Bij},
\end{equation}
The coefficients ${M}^{AB}$ in the rhs
may be considered as a $6\times 6$ matrix ${M}$, see Eq.(\ref{hom}).
Let us show  that $M\in SO_+(3,3)$ and that
Eq.(\ref{etaA}) describes an isomorphism 
$ SL(4)/(\pm {\bf 1}) \equiv SO_+(3,3)$.

The symbols $\eta^{Aij}$ and $\tilde \eta^{Aij} = 
(1/2) \epsilon^{ijkl}\eta^{Akl}$ 
satisfy relations which follow directly from
the usual relations for the 't Hooft symbols
\begin{eqnarray}\label{etao}
\eta^{Aij}\eta^{Bij} &= &\tilde \eta^{Aij} 
\tilde \eta^{Bij}=4 \delta_{AB},\\
\label{tild}
\eta^{Aij} \tilde \eta^{Bij}&=& 4\Sigma^{AB},
\end{eqnarray}
where $\Sigma$ is defined in (\ref{MSM}).
Using (\ref{etao}) 
one can present (\ref{hom1}) in the equivalent useful form
\begin{equation} \label{htra}
H^{im}H^{jn}\eta^{Amn} = {M}^{AB}\eta^{Bij}.
\end{equation}
Let us define by $\eta'^A_{ij}$ the lhs of this equation 
$\eta'^A_{ij}=
H^{im}H^{jn}\eta^{Amn}$. Then using the condition 
$\det H = 1$ 
it is easy to verify that (\ref{tild}) 
remains valid for the 
primed symbols
as well, $ \eta'^A_{ij} \tilde  \eta'^B_{ij}=4 \Sigma_{AB}$.  
Then from (\ref{htra})  one finds
 restrictions   on  the matrix ${M}$ which show that
${M}\in SO(3,3)$.
Thus (\ref{hom1}) gives a function $H\rightarrow {M}$
from $SL(4)$ into $SO(3,3)$.
From (\ref{htra}) it is easy to deduce 
that this function is a homomorphism: if
$H_1 \rightarrow {M}_1,~ H_2\rightarrow {M}_2$ 
then $ H_1 H_2 \rightarrow
{M}_2 {M}_1$.

Let us  show now that for
any ${M} \in SO_+(3,3)$ there exists $H \in SL(4)$ 
satisfying (\ref{hom1}),
while for ${M} \in SO_-(3,3)$ 
this equation cannot be satisfied. 
With this purpose consider 
representation (\ref{ccdd}) for some
${M}\in SO(3,3)$.
Notice first that if ${M}$ has the form
%?\begin{equation} \label{so4}
\begin{displaymath}
M=M_C
=\left( \begin{array}{cc} C & 0 \\ 0 & \bar C 
\end{array} \right),~~~~
C,~\bar C \in SO(3),
\end{displaymath}
then (\ref{htra}) may  certainly be 
solved with respect to  the 
matrix 
$H=H_C$ which in this case belongs to $SO(4)$.
Really, if $C = \exp (-\phi^a t^a),~\bar C = 
\exp( -\bar\phi^a t^a)$,
where $t^a$ are the generators of $SO(3)$, then 
\begin{equation} \label{M1H1}
(H_C)_{ij}=\pm (\exp( \phi^a \eta^a +\bar\phi^a 
\bar \eta^a))_{ij} \in SO(4)
\end{equation}
satisfies (\ref{htra}).
For this case  Eq.(\ref{htra}) 
describes the well-known homomorphism 
$SO(4)\rightarrow SO(3)\times SO(3)$.  

In order 
to find $H$ for any ${M}\in SO(3,3)$ it is necessary now
to consider  the second factor in the rhs of (\ref{ccdd})
%\begin{equation} \label{M0}
\begin{displaymath}
{M}_0 = \left( \begin{array}{cc} \sigma U & \bar \sigma V \\
\bar \sigma V & \sigma U \end{array} \right).
\end{displaymath}
For 
${M}={M}_0$ Eq.(\ref{hom1}) 
can be solved directly in respect to $H=H_0$. This is an easy
task because $H_0$ proves to be diagonal
\begin{equation} \label{H=H0}
H_0 = \pm \frac{1}{H_{0,44}}\left( \begin{array}{cc} 
\sigma U - 
\bar \sigma V &0  \\
0 &  H_{0,44}^2 \end{array} \right).
\end{equation}
 Condition $\det H=1$
results in $H_{0,44}^2 =\det (\sigma U- \bar \sigma V )$.
One concludes from here that
 there is a solution $H_0$ when $\sigma=+1$,  no solution 
exists
for $\sigma =-1$. One finds the matrix
$H$  presenting it in a form 
similar to (\ref{ccdd}) for ${M}$
\begin{equation} \label{HH0}
H = H_D H_0 H_C^T.
\end{equation}
A way to derive
$H_C,H_D\in SO(4)$ 
is shown in (\ref{M1H1}), and $H_0\in SL(4)$  
is found in (\ref{H=H0}) with $\sigma =+1$.
Eq.(\ref{HH0})
proves explicitly 
that (\ref{hom1}) gives a homomorphism $SL(4) \rightarrow 
{\rm on}~ SO_+(3,3)$. 
It follows from Eqs.(\ref{H=H0}),(\ref{M1H1}) 
that a kernel of this homomorphism is  $\pm {\bf 1}$.
Thus (\ref{hom1})
gives an isomorphism $ SL(4)/(\pm {\bf 1}) \equiv SO_+(3,3)$.
The 15 parameters governing $SL(4)$ matrix are:
6 parameters for each  of the two $SO(4)$ matrixes
in (\ref{HH0}) plus 3 parameters $U_i$ governing $H_0$
according to (\ref{H=H0}).

In conclusion, Eq.(\ref{hom}) is true.

\end{document}